\pdfoutput=1

\documentclass[12pt,a4paper]{article}

\usepackage{ifthen} 
\newboolean{pdflatex}
\setboolean{pdflatex}{true} 

\newboolean{articletitles}
\setboolean{articletitles}{true} 

\newboolean{uprightparticles}
\setboolean{uprightparticles}{false} 

\newboolean{inbibliography}
\setboolean{inbibliography}{false} 

\usepackage{microtype}
\usepackage{lineno}  
\usepackage{xspace} 
\usepackage{caption} 

\usepackage{graphicx}  
\usepackage{color}
\usepackage{colortbl}
\graphicspath{{./figs/}} 

\usepackage{amsmath} 
\usepackage{amssymb}
\usepackage{amsfonts}
\usepackage{upgreek} 

\newcommand*\patchAmsMathEnvironmentForLineno[1]{%
\expandafter\let\csname old#1\expandafter\endcsname\csname #1\endcsname
\expandafter\let\csname oldend#1\expandafter\endcsname\csname
end#1\endcsname
 \renewenvironment{#1}%
   {\linenomath\csname old#1\endcsname}%
   {\csname oldend#1\endcsname\endlinenomath}%
}
\newcommand*\patchBothAmsMathEnvironmentsForLineno[1]{%
  \patchAmsMathEnvironmentForLineno{#1}%
  \patchAmsMathEnvironmentForLineno{#1*}%
}
\AtBeginDocument{%
\patchBothAmsMathEnvironmentsForLineno{equation}%
\patchBothAmsMathEnvironmentsForLineno{align}%
\patchBothAmsMathEnvironmentsForLineno{flalign}%
\patchBothAmsMathEnvironmentsForLineno{alignat}%
\patchBothAmsMathEnvironmentsForLineno{gather}%
\patchBothAmsMathEnvironmentsForLineno{multline}%
\patchBothAmsMathEnvironmentsForLineno{eqnarray}%
}

\usepackage{hyperref}    
\usepackage[all]{hypcap} 

\def\lhcb {\mbox{LHCb}\xspace}

\def\babar  {\mbox{BaBar}\xspace}

\def\MagUp {\mbox{\em Mag\kern -0.05em Up}\xspace}

\ifthenelse{\boolean{uprightparticles}}%
{

 \def\Pmu         {\ensuremath{\upmu}\xspace}

 \def\Ppi         {\ensuremath{\uppi}\xspace}                 
                  
 \def\Prho        {\ensuremath{\uprho}\xspace}

 \def\PDelta      {\ensuremath{\Delta}\xspace}                 
 \def\PXi      {\ensuremath{\Xi}\xspace}                 
 \def\PLambda      {\ensuremath{\Lambda}\xspace}                 
 \def\PSigma      {\ensuremath{\Sigma}\xspace}                 
 \def\POmega      {\ensuremath{\Omega}\xspace}                 
 \def\PUpsilon      {\ensuremath{\Upsilon}\xspace}                 
 
 \def\PB      {\ensuremath{\mathrm{B}}\xspace}                 
                  
 \def\PD      {\ensuremath{\mathrm{D}}\xspace}

 \def\PK      {\ensuremath{\mathrm{K}}\xspace}

 \def\Pb      {\ensuremath{\mathrm{b}}\xspace}                 
 \def\Pc      {\ensuremath{\mathrm{c}}\xspace}

 \def\Ph      {\ensuremath{\mathrm{h}}\xspace}                 
 \def\Pi      {\ensuremath{\mathrm{i}}\xspace}

 \def\Pp      {\ensuremath{\mathrm{p}}\xspace}

 \def\Ps      {\ensuremath{\mathrm{s}}\xspace}

}
{

 \def\Pmu         {\ensuremath{\mu}\xspace}

 \def\Ppi         {\ensuremath{\pi}\xspace}                 
                  
 \def\Prho        {\ensuremath{\rho}\xspace}

 \mathchardef\PDelta="7101
 \mathchardef\PXi="7104
 \mathchardef\PLambda="7103
 \mathchardef\PSigma="7106
 \mathchardef\POmega="710A
 \mathchardef\PUpsilon="7107
                  
 \def\PB      {\ensuremath{B}\xspace}                 
                  
 \def\PD      {\ensuremath{D}\xspace}

 \def\PK      {\ensuremath{K}\xspace}

 \def\Pb      {\ensuremath{b}\xspace}                 
 \def\Pc      {\ensuremath{c}\xspace}

 \def\Ph      {\ensuremath{h}\xspace}                 
 \def\Pi      {\ensuremath{i}\xspace}

 \def\Pp      {\ensuremath{p}\xspace}

 \def\Ps      {\ensuremath{s}\xspace}

}

\makeatletter
\ifcase \@ptsize \relax
  \newcommand{\miniscule}{\@setfontsize\miniscule{4}{5}}
\or
  \newcommand{\miniscule}{\@setfontsize\miniscule{5}{6}}
\or
  \newcommand{\miniscule}{\@setfontsize\miniscule{5}{6}}
\fi
\makeatother

\DeclareRobustCommand{\optbar}[1]{\shortstack{{\miniscule (\rule[.5ex]{1.25em}{.18mm})}
  \\ [-.7ex] $#1$}}

\def\mup        {{\ensuremath{\Pmu^+}}\xspace}
\def\mun        {{\ensuremath{\Pmu^-}}\xspace} 

\def\squark    {{\ensuremath{\Ps}}\xspace}

\def\cquark    {{\ensuremath{\Pc}}\xspace}

\def\bquark    {{\ensuremath{\Pb}}\xspace}

\def\pion   {{\ensuremath{\Ppi}}\xspace}
\def\piz    {{\ensuremath{\pion^0}}\xspace}

\def\pip    {{\ensuremath{\pion^+}}\xspace}
\def\pim    {{\ensuremath{\pion^-}}\xspace}

\def\pimp   {{\ensuremath{\pion^\mp}}\xspace}

\def\rhoz     {{\ensuremath{\rhomeson^0}}\xspace}

\def\kaon    {{\ensuremath{\PK}}\xspace}
  \def\Kbar    {{\kern 0.2em\overline{\kern -0.2em \PK}{}}\xspace}

\def\KorKbar    {\kern 0.18em\optbar{\kern -0.18em K}{}\xspace}
\def\Kz      {{\ensuremath{\kaon^0}}\xspace}

\def\Kp      {{\ensuremath{\kaon^+}}\xspace}
\def\Km      {{\ensuremath{\kaon^-}}\xspace}
\def\Kpm     {{\ensuremath{\kaon^\pm}}\xspace}

\def\KS      {{\ensuremath{\kaon^0_{\rm\scriptscriptstyle S}}}\xspace}

\def\Kstarz  {{\ensuremath{\kaon^{*0}}}\xspace}

\def\Kstar   {{\ensuremath{\kaon^*}}\xspace}

  \def\Dbar    {{\kern 0.2em\overline{\kern -0.2em \PD}{}}\xspace}
\def\D       {{\ensuremath{\PD}}\xspace}

\def\DorDbar    {\kern 0.18em\optbar{\kern -0.18em D}{}\xspace}
\def\Dz      {{\ensuremath{\D^0}}\xspace}

\def\Dstarp  {{\ensuremath{\D^{*+}}}\xspace}

\def\B       {{\ensuremath{\PB}}\xspace}
\def\Bbar    {{\ensuremath{\kern 0.18em\overline{\kern -0.18em \PB}{}}}\xspace}

\def\BorBbar    {\kern 0.18em\optbar{\kern -0.18em B}{}\xspace}

\def\Bd      {{\ensuremath{\B^0}}\xspace}
\def\Bs      {{\ensuremath{\B^0_\squark}}\xspace}

  \def\Y#1S{\ensuremath{\PUpsilon{(#1S)}}\xspace}

\def\proton      {{\ensuremath{\Pp}}\xspace}

\def\Lbar        {{\ensuremath{\kern 0.1em\overline{\kern -0.1em\PLambda}}}\xspace}
\def\LorLbar    {\kern 0.18em\optbar{\kern -0.18em \PLambda}{}\xspace}

\def\BF         {{\ensuremath{\cal B}}\xspace}

\newcommand{\decay}[2]{\ensuremath{#1\!\to #2}\xspace}         

\def\to                 {\ensuremath{\rightarrow}\xspace}

\def\CP                {{\ensuremath{C\!P}}\xspace}

\def\AT#1     {\ensuremath{A_{\mathrm{T}}^{#1}}\xspace}           

\def\C#1      {\ensuremath{\mathcal{C}_{#1}}\xspace}                       
\def\Cp#1     {\ensuremath{\mathcal{C}_{#1}^{'}}\xspace}                    
\def\Ceff#1   {\ensuremath{\mathcal{C}_{#1}^{\mathrm{(eff)}}}\xspace}        
\def\Cpeff#1  {\ensuremath{\mathcal{C}_{#1}^{'\mathrm{(eff)}}}\xspace}       
\def\Ope#1    {\ensuremath{\mathcal{O}_{#1}}\xspace}                       
\def\Opep#1   {\ensuremath{\mathcal{O}_{#1}^{'}}\xspace}                    

\newcommand{\tev}{\ifthenelse{\boolean{inbibliography}}{\ensuremath{~T\kern -0.05em eV}\xspace}{\ensuremath{\mathrm{\,Te\kern -0.1em V}}}\xspace}
\newcommand{\gev}{\ensuremath{\mathrm{\,Ge\kern -0.1em V}}\xspace}
\newcommand{\mev}{\ensuremath{\mathrm{\,Me\kern -0.1em V}}\xspace}
\newcommand{\kev}{\ensuremath{\mathrm{\,ke\kern -0.1em V}}\xspace}
\newcommand{\ev}{\ensuremath{\mathrm{\,e\kern -0.1em V}}\xspace}
\newcommand{\gevc}{\ensuremath{{\mathrm{\,Ge\kern -0.1em V\!/}c}}\xspace}
\newcommand{\mevc}{\ensuremath{{\mathrm{\,Me\kern -0.1em V\!/}c}}\xspace}
\newcommand{\gevcc}{\ensuremath{{\mathrm{\,Ge\kern -0.1em V\!/}c^2}}\xspace}
\newcommand{\gevgevcccc}{\ensuremath{{\mathrm{\,Ge\kern -0.1em V^2\!/}c^4}}\xspace}
\newcommand{\mevcc}{\ensuremath{{\mathrm{\,Me\kern -0.1em V\!/}c^2}}\xspace}

\def\mum  {\ensuremath{{\,\upmu\rm m}}\xspace}

\def\invfb   {\ensuremath{\mbox{\,fb}^{-1}}\xspace}

\def\gsim{{~\raise.15em\hbox{$>$}\kern-.85em
          \lower.35em\hbox{$\sim$}~}\xspace}
\def\lsim{{~\raise.15em\hbox{$<$}\kern-.85em
          \lower.35em\hbox{$\sim$}~}\xspace}

\def\ptot       {\mbox{$p$}\xspace}
\def\pt         {\mbox{$p_{\rm T}$}\xspace}

\def\evtgen     {\mbox{\textsc{EvtGen}}\xspace}

\def\geant      {\mbox{\textsc{Geant4}}\xspace}

\def\photos     {\mbox{\textsc{Photos}}\xspace}

\def\pythia     {\mbox{\textsc{Pythia}}\xspace}

\def\tell1  {TELL1\xspace}
\def\ukl1   {UKL1\xspace}

\newcommand{\eg}{\mbox{\itshape e.g.}\xspace}

\ifthenelse{\boolean{uprightparticles}}%
{\def\Prho      {\ensuremath{\uprho}\xspace}
 
}
{\def\Prho      {\ensuremath{\rho}\xspace}
 
}
\def\rhoz   {\ensuremath{\Prho^0}\xspace}

\def\had  {\ensuremath{\Ph}\xspace}

\def\hadpm  {\ensuremath{\Ph^{\pm}}\xspace}

\def\hadprimmp  {\ensuremath{\had^{'\mp}}\xspace}

\def\BdtoKzPiPi   {\decay{\Bd}{\Kz \pip \pim}}

\def\BdtoKsPiPi   {\decay{\Bd}{\KS \pip \pim}}

\def\KsPiPi{\ensuremath{\KS \pip \pim}\xspace}
\def\KsKPi{\ensuremath{\KS \Kpm \pimp}\xspace}

\def\Ks{\ensuremath{\KS}\xspace}
\def\Kstz  {\ensuremath{K^{*0}}\xspace}

\def\BstoKstarKs {\decay{\Bs}{\KS \Kstz}}
\def\BdtoKstarKs {\decay{\Bd}{\KS \Kstz}}

\usepackage{cite} 
\usepackage{mciteplus}

\usepackage{longtable} 

\newcommand{\myBdstoKstarKs}{\decay{\B_{(s)}^0}{\KS \Kstz}}
\newcommand{\myBdstoKshhp}{\decay{\B_{(s)}^0}{\KS \hadpm \hadprimmp}}
\newcommand{\myBdstoKsKPi}{\decay{\B_{(s)}^0}{\KS \Kpm \pimp}}

\date{4th January 2016}
\begin{document}

\renewcommand{\thefootnote}{\fnsymbol{footnote}}
\setcounter{footnote}{1}


\begin{titlepage}
\pagenumbering{roman}

\vspace*{-1.5cm}
\centerline{\large EUROPEAN ORGANIZATION FOR NUCLEAR RESEARCH (CERN)}
\vspace*{1.5cm}
\hspace*{-0.5cm}
\begin{tabular*}{\linewidth}{lc@{\extracolsep{\fill}}r}
\ifthenelse{\boolean{pdflatex}}
{\vspace*{-2.7cm}\mbox{\!\!\!\includegraphics[width=.14\textwidth]{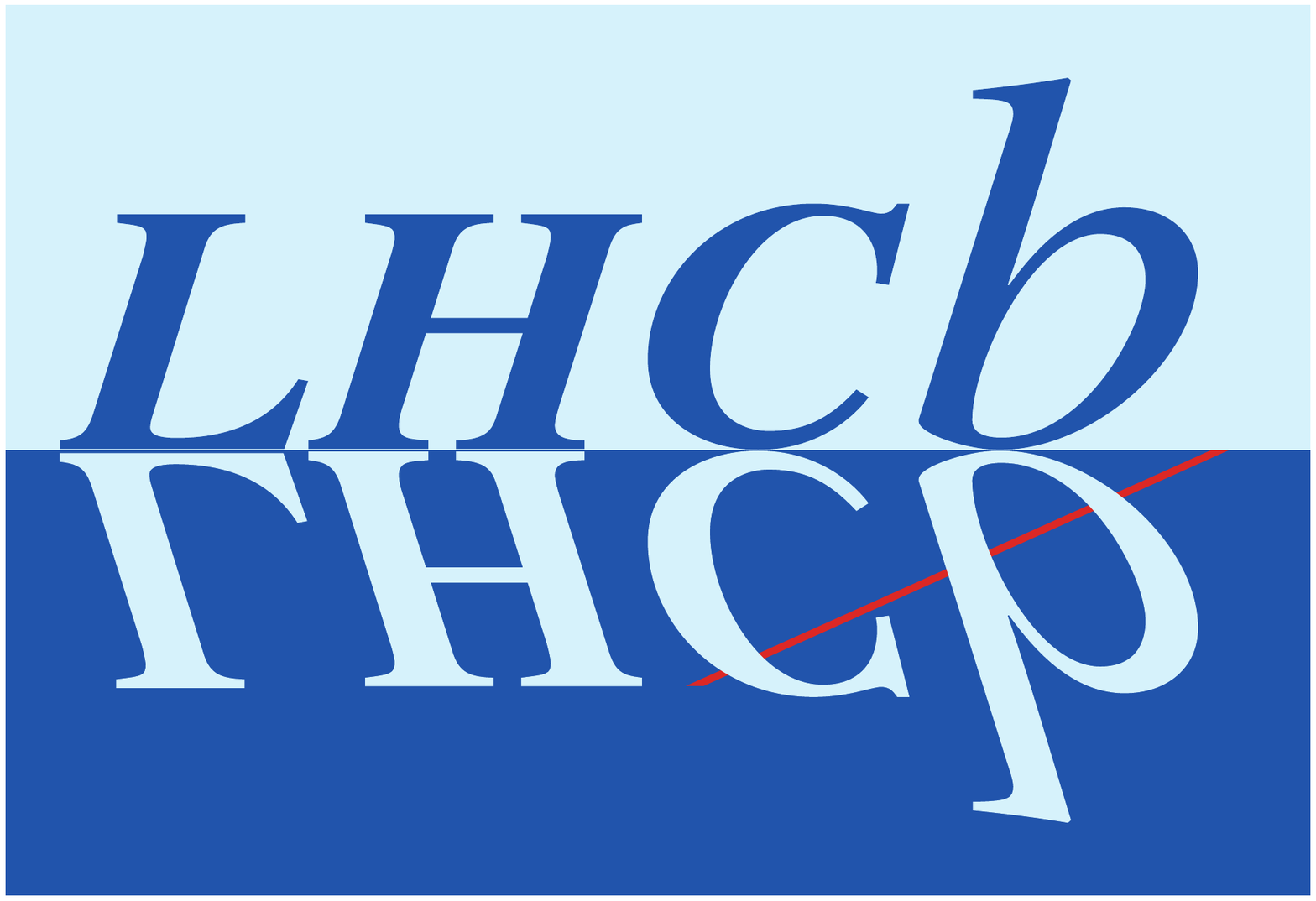}} & &}%
{\vspace*{-1.2cm}\mbox{\!\!\!\includegraphics[width=.12\textwidth]{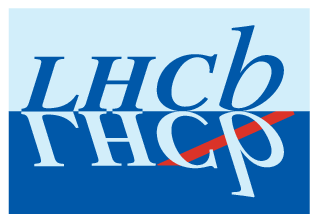}} & &}%
\\
 & & CERN-PH-EP-2015-144 \\  
 & & LHCb-PAPER-2015-018 \\  
 & & 4 January 2016 \\ 
 & & \\
\end{tabular*}

\vspace*{2.0cm}

{\bf\boldmath\huge
\begin{center}
First observation of the decay $B_{s}^{0} \to K_{\rm{S}}^{0} K^{*}(892)^{0}$
\end{center}
}

\vspace*{2.0cm}

\begin{center}
The LHCb collaboration\footnote{Authors are listed at the end of this paper.}
\end{center}

\vspace{\fill}

\begin{abstract}
  \noindent
A search for $B_{(s)}^{0} \to K_{S}^{0} K^{*}(892)^{0}$ decays is performed using $pp$ collision data, corresponding 
to an integrated luminosity of $1.0~\text{fb}^{-1}$, collected with the LHCb detector at a centre-of-mass energy of $7~\text{TeV}$. 
The $B_{s}^{0} \to K_{S}^{0} K^{*}(892)^{0}$ decay is observed for the first time, with a significance of 7.1 standard deviations.
The branching fraction is measured to be 
\begin{equation*}
\mathcal{B}(B_{s}^{0} \to \bar{K}^{0} K^{*}(892)^{0}) + \mathcal{B}(B_{s}^{0} \to K^{0} \bar{K}^{*}(892)^{0}) =  (16.4 \pm 3.4 \pm 2.3) \times 10^{-6},\\
\end{equation*}
where the first uncertainty is statistical and the second is systematic.
No evidence is found for the decay $B^{0} \to K_{S}^{0} K^{*}(892)^{0}$ and an upper limit is set on the branching fraction,
$\mathcal{B}(B^{0} \to \bar{K}^{0} K^{*}(892)^{0}) + \mathcal{B}(B^{0} \to K^{0} \bar{K}^{*}(892)^{0}) < 0.96 \ \times 10^{-6}, $ at $90\,\% $ confidence level. 
All results are consistent with Standard Model predictions.

\end{abstract}


\begin{center}
Published in JHEP 2016(1), 1-17
\end{center}

\vspace{\fill}

{\footnotesize 
\centerline{\copyright~CERN on behalf of the \lhcb collaboration, licence \href{http://creativecommons.org/licenses/by/4.0/}{CC-BY-4.0}.}}
\vspace*{2mm}

\end{titlepage}


\newpage
\setcounter{page}{2}
\mbox{~}
%
%
%
%

\cleardoublepage


\renewcommand{\thefootnote}{\arabic{footnote}}
\setcounter{footnote}{0}



\pagestyle{plain} 
\setcounter{page}{1}
\pagenumbering{arabic}


%

\section{Introduction}
\label{sec:Introduction}

Violation of the combined charge-conjugation and parity symmetry (\CP) 
is one of the fundamental ingredients to explain a dynamical generation of 
the observed matter-antimatter asymmetry in the universe\cite{sakharov}. In the Standard Model of particle physics (SM),
\CP violation in the quark sector is generated by a single complex phase in the Cabibbo-Kobayashi-Maskawa 
matrix~\cite{Cabibbo,Kobayashi}. However, the observed baryon asymmetry is too large to be explained by the SM mechanism alone~\cite{Riotto}. 
Non-leptonic \B meson decays dominated by amplitudes involving a quark and a $W$ boson in a loop (penguin)
are sensitive to the presence of non-SM physics processes. 
These processes could provide additional sources of \CP violation that might explain the observed baryon asymmetry.
The \myBdstoKshhp ($h, h^{'}=\pi,K$)
decays are interesting for \CP violation measurements\footnote{Charge-conjugate modes are implicitly included throughout this paper.}.
Knowledge of the branching fractions of the various sub-modes, as reported
in this paper, is an important input to the theory of \CP-violation, particularly
models of new-physics contributions to $b \to s$ transitions Ref.~\cite{Ciuchini} . The
measurements also allow tests of QCD models (see, for example, the
predictions in Refs. ~\cite{Cheng:2009mu,Ali:2007ff,Su:2011eq}).
If sufficient data are available, a common approach for three-body decays 
is to perform an amplitude analysis by studying the structure of the Dalitz plot~\cite{Dalitz:1953cp}. 
If data are less abundant and the decay products originate from intermediate resonances, as in the present analysis, a quasi two-body approach can be used.

The LHCb collaboration has provided results for inclusive \myBdstoKshhp decays~\cite{LHCb-PAPER-2013-042}, 
and more recently the first measurements of $B_s^0$ meson decays to $K^{*}(892)^{-}\pi^{+}$ and $K^{*}(892)^{-} K^{+}$ final
states~\cite{LHCb-PAPER-2014-043}. 
An initial search for the neutral decay
$\Bd \to K^{*}(892)^{0}\Ks$ has been reported by the \babar experiment~\cite{Aubert:2006wu}. 
In this paper a search for \decay{\B_{(s)}^0}{\KS \Kstar(892)^{0}} decays is reported, where the $\Kstar(892)^{0}$ meson,
hereafter denoted by \Kstarz, decays to the \Kp\pim final state. 
The resonant structure in the \Kp\pim invariant mass region around 1\gevcc
is analysed to determine the number of decays that proceed through an
intermediate \Kstarz resonance.
The branching fraction is measured relative to the \BdtoKsPiPi decay~\cite{PDG2014}, using the relation
\begin{equation}
       \label{eq : master-formula}
        \frac{\mathcal{B}(\myBdstoKstarKs)}{\mathcal{B}(\BdtoKsPiPi)} = \frac{N_{\myBdstoKstarKs}}{N_{\BdtoKsPiPi}} \cdot \frac{\epsilon_{\BdtoKsPiPi}}{\epsilon_{\myBdstoKstarKs}} \cdot 
        \frac{f_{d}}{f_{d(s)}} \cdot  \frac{1}{\mathcal{B}(K^{*0} \to K^+ \pi^-)},
\end{equation}
where $N$ represents the number of observed decays, $\epsilon$ the total efficiency,
and $f_s/f_d$ the ratio of the fragmentation fractions of a $b$ quark into a \Bs or a \Bd meson~\cite{LHCb-PAPER-2011-018,LHCb-PAPER-2012-037,LHCb-CONF-2013-011} and $\mathcal{B}(K^{*0} \to K^+ \pi^-)$ is the branching fraction of the $K^{*0}$ meson into $K^+ \pi^-$ final state.
In the following, the \myBdstoKstarKs and \BdtoKsPiPi decays are referred to as signal and normalisation channels, respectively.

\section{Detector and simulation}
\label{sec:Detector}
The analysis is performed using $pp$ collision data recorded with the LHCb detector, 
corresponding to an integrated luminosity of $1.0$\invfb, at a centre-of-mass energy of $7$\tev.
The \lhcb detector~\cite{Alves:2008zz,LHCb-DP-2014-002} is a single-arm forward
spectrometer covering the \mbox{pseudorapidity} range $2<\eta <5$,
designed for the study of particles containing \bquark or \cquark
quarks. The detector includes a high-precision tracking system
consisting of a silicon-strip vertex detector surrounding the $pp$
interaction region, a large-area silicon-strip detector located
upstream of a dipole magnet with a bending power of about
$4{\rm\,Tm}$, and three stations of silicon-strip detectors and straw
drift tubes placed downstream of the magnet.
The tracking system provides a measurement of momentum, \ptot, of charged particles with
a relative uncertainty that varies from 0.5\% at low momentum to 1.0\% at 200\gevc.
The minimum distance of a track to a primary vertex~(PV), the impact parameter, is measured with a resolution of $(15+29/\pt)\mum$,
where \pt is the component of the momentum transverse to the beam, in \gevc.
Different types of charged hadrons are distinguished using information
from two ring-imaging Cherenkov detectors. 
Photons, electrons and hadrons are identified by a calorimeter system consisting of
scintillating-pad and preshower detectors, an electromagnetic
calorimeter and a hadronic calorimeter. Muons are identified by a
system composed of alternating layers of iron and multiwire
proportional chambers.

Simulated events are used to determine the efficiency of the selection requirements,
to study possible sources of background and to determine the parametrisations used to model the data.
In the simulation, $pp$ collisions are generated using
\pythia6~\cite{Sjostrand:2006za} with a specific \lhcb
configuration~\cite{LHCb-PROC-2010-056}.  Decays of hadronic particles
are described by \evtgen~\cite{Lange:2001uf}, in which final-state
radiation is generated using \photos~\cite{Golonka:2005pn}. The
interaction of the generated particles with the detector, and its response,
are implemented using the \geant toolkit~\cite{Allison:2006ve, *Agostinelli:2002hh}
as described in Ref.~\cite{LHCb-PROC-2011-006}.

\section{Event selection}
\label{sec:selection}

The online event selection system (trigger)~\cite{LHCb-DP-2012-004} consists of a
hardware stage, based on information from the calorimeter and muon
systems, followed by a software stage, in which all charged particles
with $\pt>500\mevc$ are reconstructed. 
The hardware hadron trigger requires a calorimeter cluster with transverse energy greater than $3.5\gev$.
In the offline selection, candidates are divided into two non mutually exclusive categories based on the hardware
trigger decision.  
One category consists of candidates whose decay products satisfy the hadron trigger requirements, 
while the other consists of candidates from events in which other particles meet the hardware trigger requirements.
Only events that fall into either of these categories are used in the subsequent
analysis.
The software trigger requires a two-, three- or four-particle
secondary vertex with a significant displacement from the primary
$pp$ interaction points. At least one charged particle
must have $\pt > 1.7\gevc$ and be
inconsistent with originating from any PV.
A multivariate algorithm~\cite{BBDT} is used for
the identification of secondary vertices consistent with the decay
of a \bquark hadron.

In the offline selection the \myBdstoKstarKs decays are reconstructed through the \Kstarz\to\Kp\pim and \KS\to\pip\pim decay modes,
where the \KS candidate is constrained to its known mass~\cite{PDG2014} and the \B candidate is constrained to originate from a PV.
Decays of \KS mesons are reconstructed in two mutually exclusive categories:  {\it long} \KS candidates, which decay sufficiently early that their daughter pions are reconstructed in the vertex detector; 
and {\it downstream} \KS candidates, which have daughter particles that are only reconstructed in the rest of the tracking system. 
As these two categories have different backgrounds, and the long \KS mesons have better momentum and vertex resolutions, the selection criteria for long and downstream \KS candidates differ.
The selection criteria follow those in Ref.~\cite{LHCb-PAPER-2013-042}.

Fully reconstructed background decays that have the same final state as the signal
include contributions from \B decays to final states involving charm mesons, such as
$D h$, with a $\KS h^+ h^-$ final state, or $\Lambda_b^0$ decays to $\Lambda_c^+h^-$, with \decay{\Lambda_c^+}{\KS p}, where the proton is misidentified as a \pip or \Kp.
In addition, \B decays with an intermediate charmonium state like \Bd\to$J/\psi$\KS, with \decay{J/\psi}{\pip\pim,\Kp\Km,\mup\mun}, 
can be present in the mass region of the normalisation channel. To reduce the contamination from
these backgrounds, a veto is applied on the invariant mass of each of the possible intermediate states reconstructed under the corresponding
hypothesis.
Candidates are excluded if the reconstructed mass of a two-body intermediate state is within 30\mevcc (48\mevcc) of the known mass 
of the relevant intermediate charm (charmonium) resonance~\cite{PDG2014} of one of the backgrounds considered.
No particle identification information is used at this stage.

If a final-state hadron is misidentified, signal yields can potentially be affected by decays 
into any $\KS h^\pm h'^{\mp}$ final state, especially when the $h^\pm h'^{\mp}$ proceeds through a resonance.
Particle identification requirements on the two tracks originating from the \B decay vertex
are used to separate pions, kaons and protons,
and to reduce this background to a negligible level.
The largest source of background is due to random tracks that form candidate \B or \KS decay vertices.
A multivariate discriminant based on a boosted decision tree (BDT) algorithm~\cite{Breiman,AdaBoost} is used to reduce
this background. 
The greatest discrimination in the BDT is provided by kinematic properties of the \B meson, its flight direction with respect 
to the PV, and variables defined analogously for its decay products.
The optimisation of the BDT is described in Ref.~\cite{LHCb-PAPER-2013-042};
the selection requirement on the BDT response for this analysis is chosen to maximise 
$\epsilon / (a/2 + \sqrt{N_\B})$~\cite{Punzi:2003bu}.
Here, $\epsilon$ is the signal efficiency, 
\B represents the number of background events in the signal mass interval,
which is estimated using data by extrapolating the number of background events
from the upper mass sideband into the signal region, and $a=5$ is the chosen target signal significance. 

The efficiencies are determined from simulation, except for
the particle identification
efficiencies. The latter are determined from data using samples of kinematically identified charged particles from \decay{\Dstarp}{\Dz\pip} with \decay{\Dz}{\Km\pip}, and \decay{\Lambda}{\pim \proton} decays, reweighted to match the kinematic properties of the signal.
The BDT selection efficiency for signal is approximately 85\% (90\%) for downstream (long) signal decays; approximately 88\% (95\%) of backgrounds in the respective categories are rejected.
The \BdtoKsPiPi decay selection efficiency is taken from Ref.~\cite{LHCb-PAPER-2013-042}.
The efficiencies for the normalisation channel are determined in bins of the Dalitz plane and are reweighted
from data using the \textit{sPlot} method~\cite{Pivk:2004ty}.

\section{Fit model}
\label{sec:fit}

Two-dimensional extended maximum likelihood fits to the unbinned \Ks\Kp\pim and \Kp\pim mass distributions are used to determine the event yields
for the signal channel, while an independent one-dimensional fit to the \Ks\pip\pim mass distribution is used for the normalisation channel. The correlation between the two signal mass distributions is checked on simulation. The results
do not show significant correlations and therefore the correlation terms are neglected in the fit. Candidates in the long and downstream categories are fitted simultaneously. 
The signal fit is restricted to candidates in the mass regions $5000$ $<$ $m$(\Ks\Kp\pim) $<$ $5800$ \mevcc and $650$ $<$ $m$(\Kp\pim) $<$ $1200$ \mevcc.
The fit model consists of signal, non-resonant background,
partially reconstructed background and combinatorial background components.

The \Bd and the \Bs components of the signal are both parametrised as
two Gaussian distributions with a power-law tail on each side.
For each component the two functions share the peak position and the width parameters.
The parameters describing the tails are determined by fits to simulated samples
and subsequently fixed in the fit to data. The systematic uncertainty associated with this choice is found to be negligible. The \Bd-\Bs mass difference is fixed to the known value\cite{PDG2014}.
The \Kstarz mass distribution is parametrised by a relativistic Breit-Wigner function with the peak position and width
allowed to vary in the fit.

The components in the \B mass model that are non-resonant in \Kp\pim are parametrised by the same function as the signal, 
sharing their peak positions and widths with the signal functions. 
The tail parameters are fixed according to the values obtained from simulation.
The non-resonant component of the \Kp\pim mass distribution is approximated by a normalised linear function as in Ref.\cite{LHCb-PAPER-2014-043}, with the zero point of the function
on the abscissa determined by the fit. 
While the ratio between the non-resonant and the signal components is fixed to be the same for the two \KS meson categories,
it is independent for the \Bd and the \Bs candidates.

Backgrounds from partially reconstructed decays are classified into two categories. Decays such as $\B\to\D h$ are parametrised
by means of ARGUS functions convolved with a Gaussian function in the \B candidate mass, and linear functions
in the \Kstarz candidate mass. The choice is based on simulation studies and previous findings~\cite{LHCb-PAPER-2013-042,LHCb-PAPER-2014-043}. In decays such as \decay{\Bs}{\Kstarz\overline{K}^{*0}}, where one resonance decays as 
\decay{\Kstarz}{\Kp\pim} while the other decays as \decay{\overline{K}^{*0}}{\KS\piz}, the \Bs mass
distribution is described using the same parametrisation as for the previous background, while the invariant mass distribution for \Kstarz candidates is described
by a relativistic Breit-Wigner function sharing the peak position and widths with the signal component. The yield for these components
are determined in the fit to data.

The combinatorial background is modelled by an exponential function in the \B candidate mass distribution 
and a linear function in the \Kstarz candidate mass distribution. These functions are found to give good agreement with the distributions in the appropriate data sidebands. The slopes of the exponential functions are independent for the long and downstream categories, while the abscissae of the linear functions are the same.
All these parameters are allowed to vary in the fit.

The parametrisation used to model the \BdtoKsPiPi normalisation and the background follow those used to fit the signal mode.
In addition two other categories of partially reconstructed backgrounds are included: decays such as $\Bd\to\KS\pip\pim\gamma$ or $\Bd\to\eta'\KS$, with \decay{\eta}{\rhoz\gamma}; and misidentified \myBdstoKsKPi decays.
Their parameters are fixed in the fit to the values derived from simulated samples.

\begin{figure}[!t ]
\begin{center}
\includegraphics[width=0.49\textwidth]{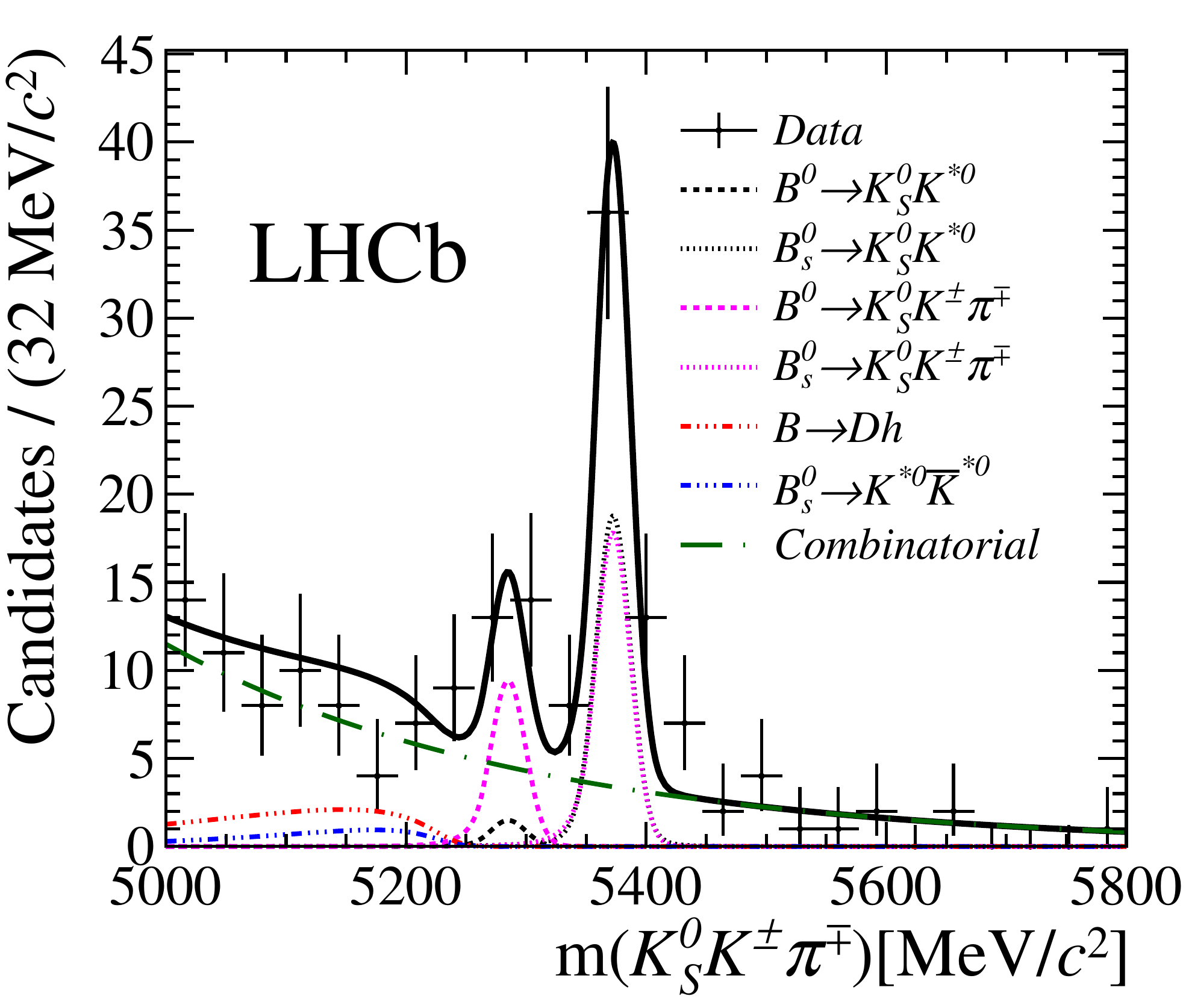}
\includegraphics[width=0.49\textwidth]{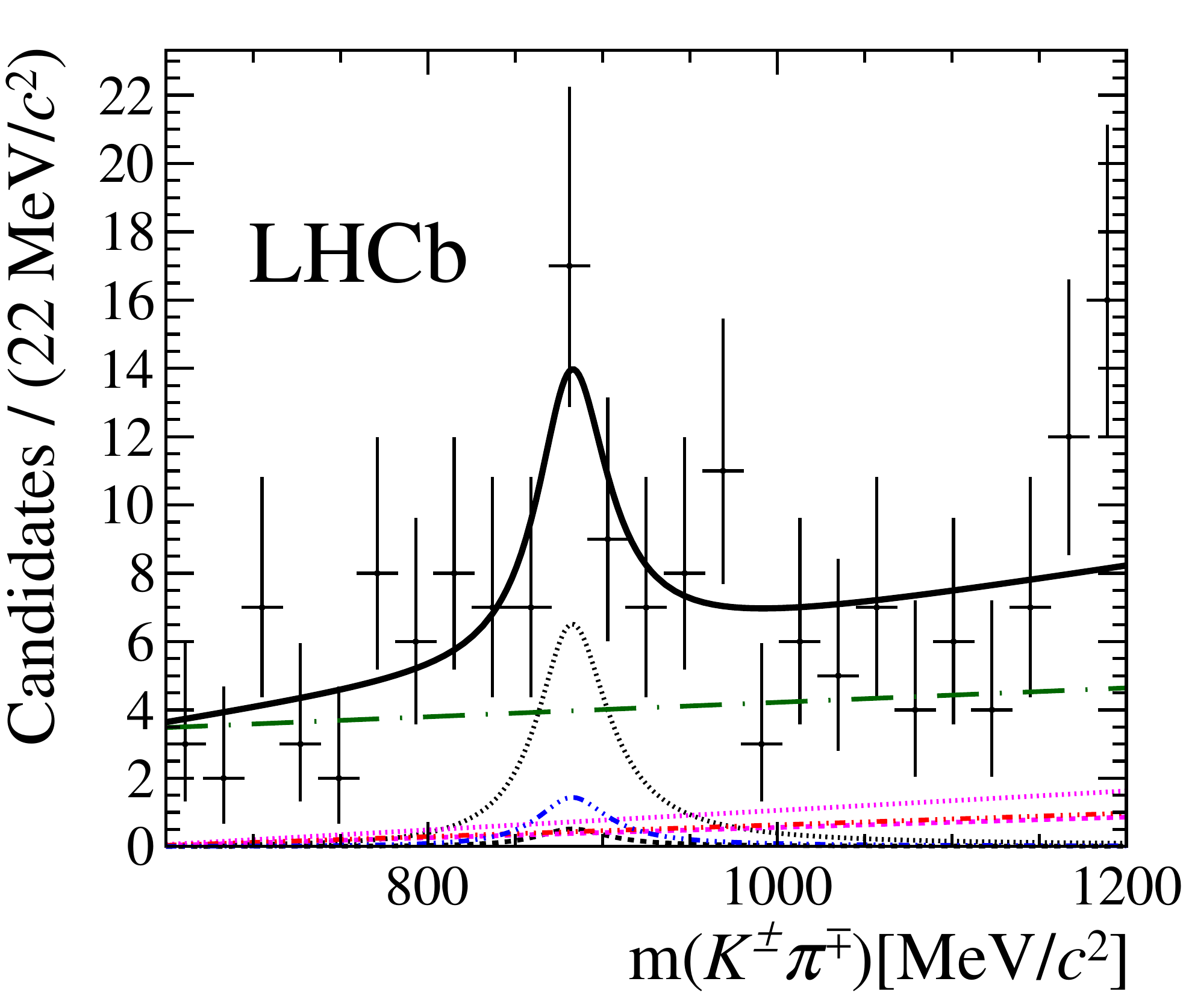}
\includegraphics[width=0.49\textwidth]{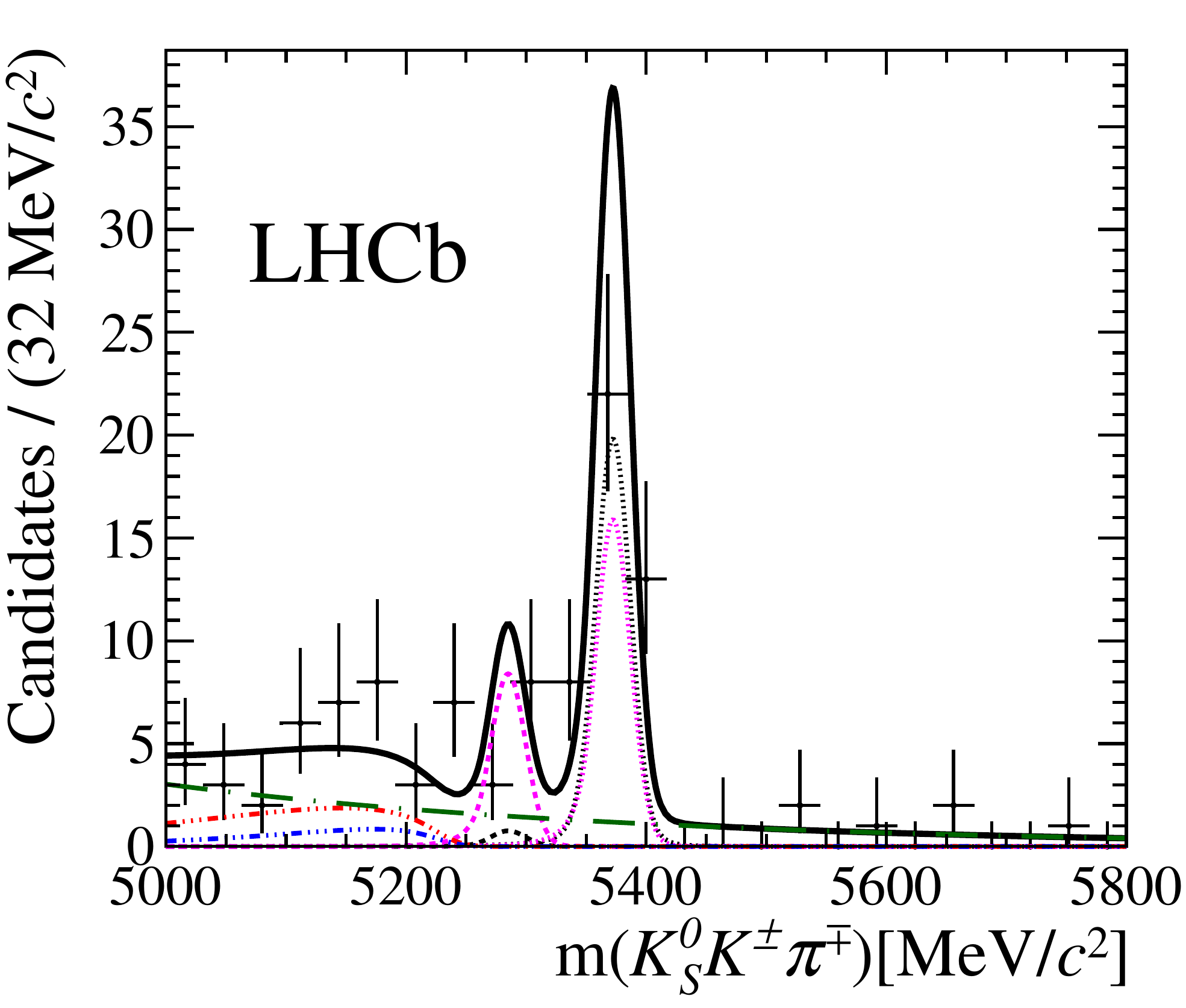}
\includegraphics[width=0.49\textwidth]{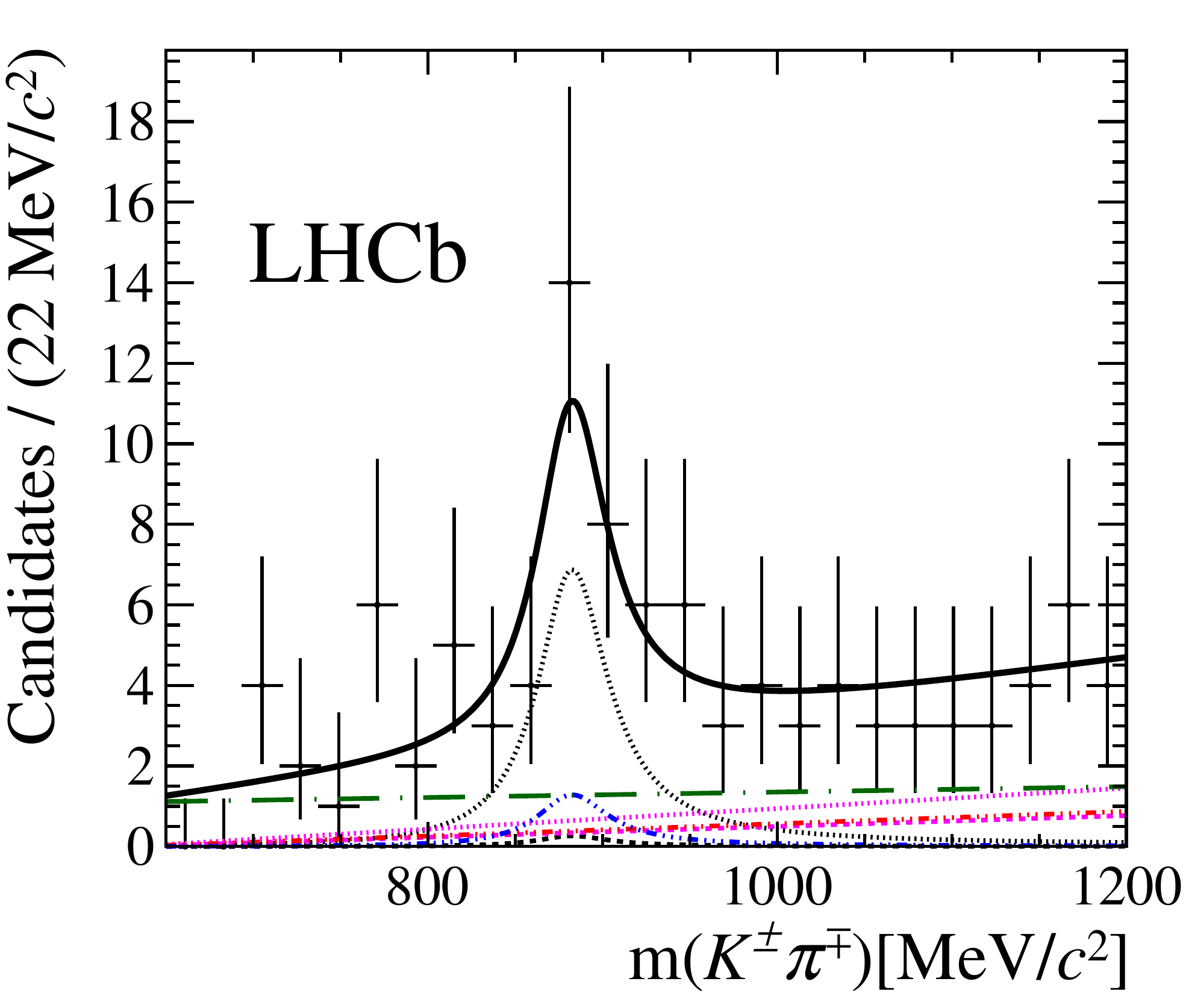}
\caption{\small
Distribution of (left) \KsKPi mass and (right)  \Kpm{}\pimp mass for signal candidates with fit results overlaid
for (top) downstream and (bottom) long categories.
The data are shown as black points with error bars. The overall fit is represented by the solid black line.
The \Bd and \Bs signal components are the black short-dashed and dotted lines respectively, while the non-resonant components are the magenta short-dashed and dotted lines.
The partially reconstructed backgrounds are the red triple-dotted line ($\B \to Dh$) and the blue triple-dotted line (\decay{\Bs}{\Kstarz\overline{K}^{*0}}).
The combinatorial background is the green long-dash dotted line.
}
\label{fig : fit-result-Kpi}
\end{center}
\end{figure}

\begin{figure}[!t ]
\begin{center}
\includegraphics[width=0.49\textwidth]{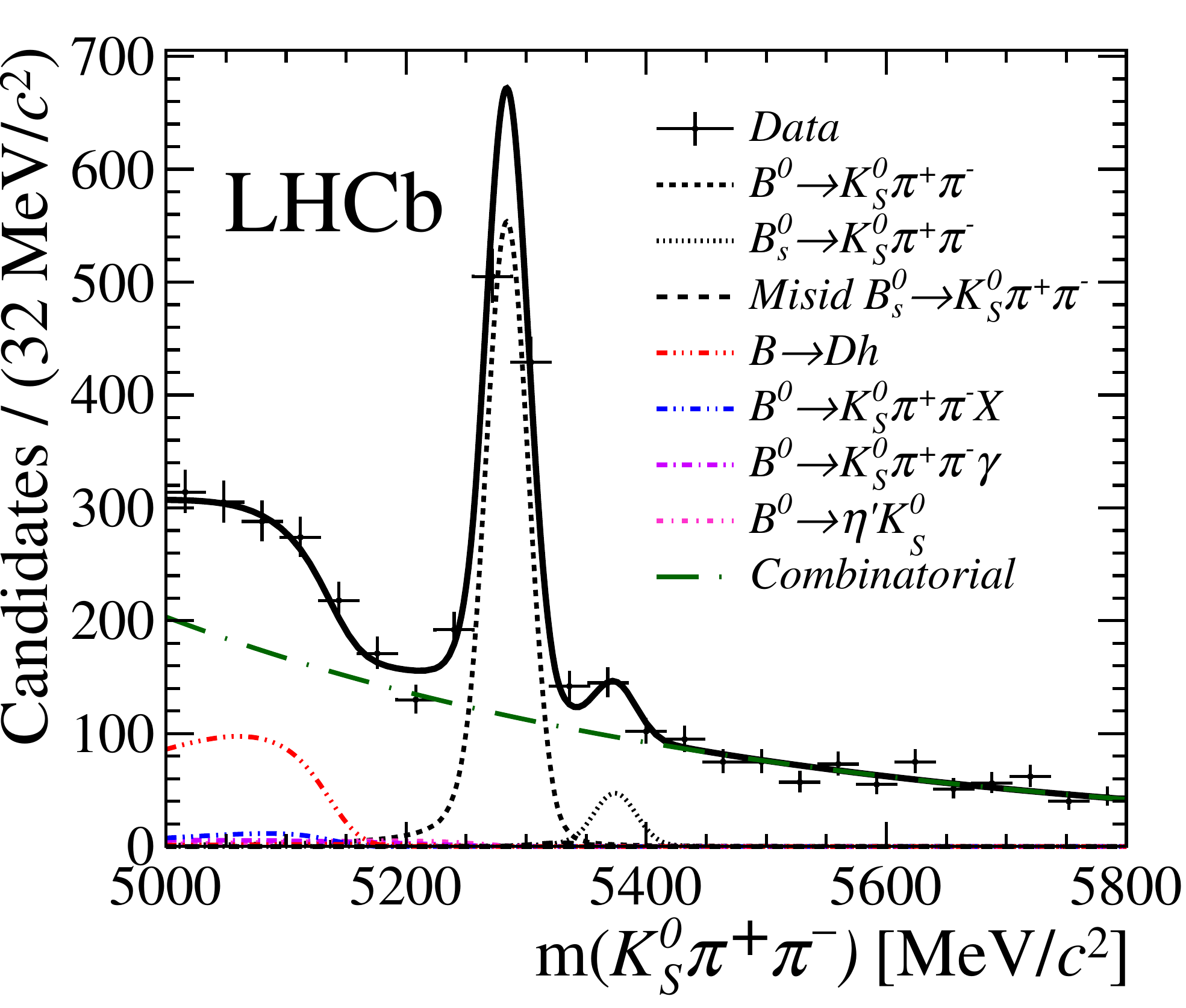}
\includegraphics[width=0.49\textwidth]{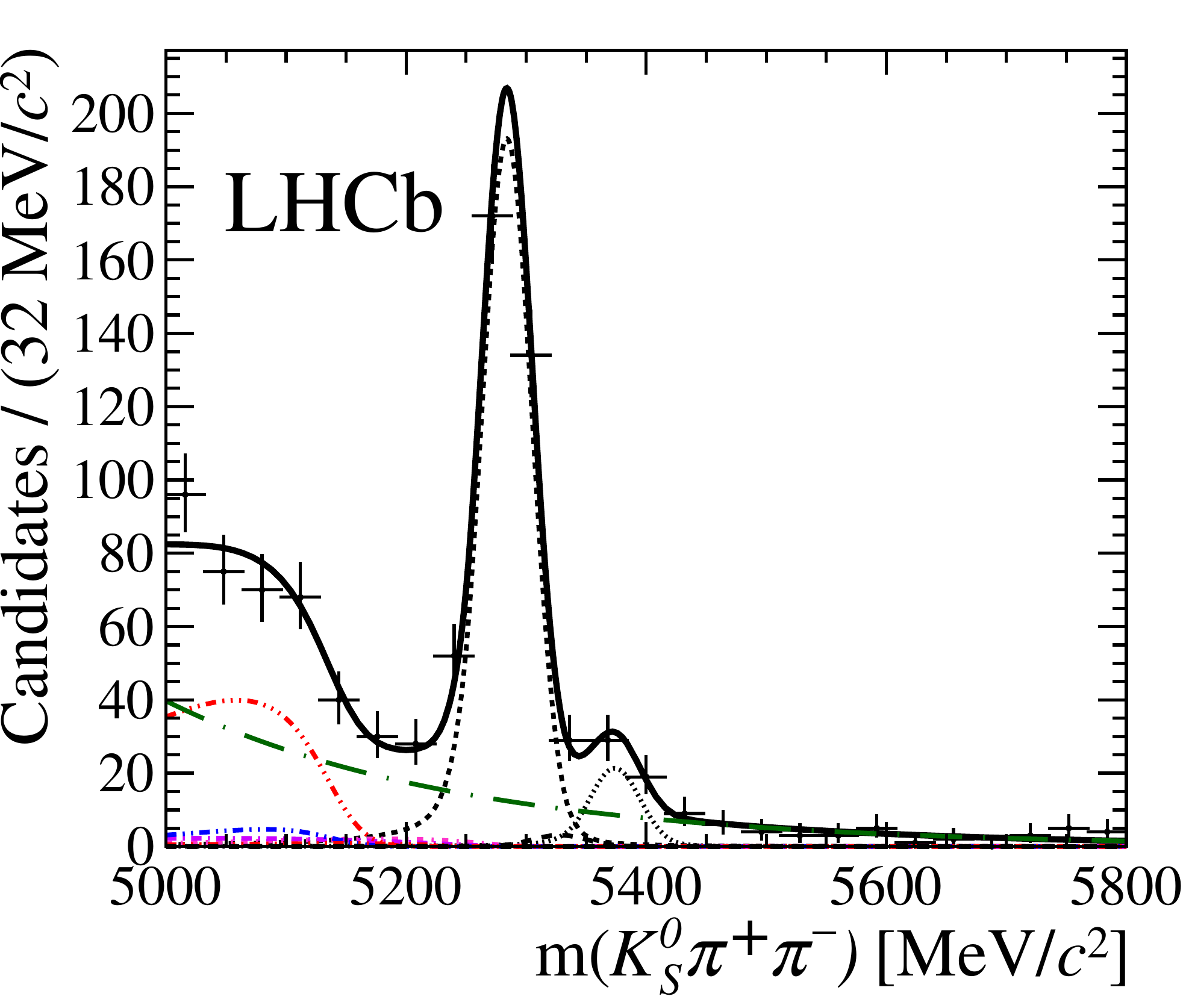}
\caption{ \small
Distribution of \KsPiPi  mass for signal candidates with fit results overlaid
for (left) downstream and (right) long categories.
The data are described by the black points with error bars. The overall fit is represented by the solid black line.
The \Bd and \Bs signal components are the black short-dashed and dotted lines.
The misidentified \Bs decay is the black dashed line, respectively.
The partially reconstructed backgrounds are the red triple-dotted line ($\B \to Dh$), the blue triple-dotted line ($\Bd \to \KS \pip \pim X$), the violet
dash single-dotted line ($\Bd\to\eta'\KS$) and the pink short-dash single dotted line ($\Bd\to\KS\pip\pim\gamma$).
The combinatorial background is the green long-dash dotted line. Some of the contributions are small in the figures.
}
\label{fig : fit-result-pipi}
\end{center}
\end{figure}

The observed \KS\Kstarz and \KS\pip\pim mass distributions and the corresponding fits are shown in Figs.~\ref{fig : fit-result-Kpi} and \ref{fig : fit-result-pipi}, respectively.
The signal yields are reported in Table~\ref{fit_results}. 
The \Bd mode is dominated by the non-resonant component. 
The statistical significance of the \Bs signal is determined using Wilks' theorem~\cite{Wilks:1938dza} 
and by combining the long and the downstream samples.
The significance including relevant systematic uncertainties, estimated by repeating the procedure with the signal likelihood convolved with a Gaussian
function of width equal to the sum in quadrature of the systematic uncertainties, is $7.1$ standard deviations.

\begin{table}[t]
    \caption{\small
Signal yields obtained from the fits to \KsKPi and \KsPiPi mass distributions and corresponding efficiencies. Only statistical contributions to the uncertainty are reported.
    }
\label{fit_results}
\begin{center}
\begin{tabular}{l c c c c}
\hline
Decay  & \multicolumn{2}{c}{Downstream} & \multicolumn{2}{c}{Long} \\
\cline{2-5}
& Yield & Efficiency ($\%$) & Yield & Efficiency ($\%$)  \\
\hline
\BstoKstarKs  &$\phantom{0}21\pm\phantom{0}6$ &$0.0174\pm0.0012$  &$\phantom{0}25\pm\phantom{0}6$ &$0.0121\pm0.0008$ \\
\BdtoKstarKs   &$\phantom{00}2\pm\phantom{0}3$ &$0.0183\pm0.0013$ &$\phantom{00}1\pm\phantom{0}2$ &$0.0125\pm0.0009$  \\
\BdtoKsPiPi  &$828\pm41$ & $0.0336\pm0.0010$ & $341\pm23$ & $0.0117\pm0.0009$  \\
\hline
\end{tabular}
\end{center}
\end{table}

\section{Systematic uncertainties}
\label{sec:systematics}
The model used to fit data and the limited knowledge in the efficiency determination are possible sources of systematic uncertainty.
Many parameters in the fit are fixed to values obtained from fits to simulated data.
The associated systematic uncertainties are determined from fits to pseudoexperiments generated assuming alternative values of the relevant parameters, corresponding to variations within uncertainties around their default values.
The average difference between the yields determined in the pseudoexperiments and
the nominal value is taken as a systematic uncertainty.

The fit model does not account for the possible interference between the P wave of the $K^{*}(892)^0$ resonance
and the S wave from other intermediate states, \eg the non-resonant component or the $K^{*}(1430)^0$ resonance.
The associated systematic uncertainty is determined by exploiting the distribution of $\theta_{\Kstarz}$, defined as 
the angle between the flight direction of the $K^+$ in the \Kstarz rest frame with respect to the direction of the boost from 
the laboratory frame to the \Kstarz rest frame.
The $\cos\theta_{\Kstarz}$ distribution is described by a parabola, where the second-order term represents
the signal P wave, the constant term is related to the S wave and the first-order term accounts for the interference.
Using the \textit{sPlot} technique~\cite{Pivk:2004ty}, the $\cos\theta_{\Kstarz}$ distribution of the signal P and S wave is unfolded
from the other background components. A fit in the region of positive $\cos\theta_{\Kstarz}$ is performed using a second-order polynomial and the systematic
uncertainty is determined as the relative difference between the integral of the function when the constant
coefficient is allowed to vary and when this coefficient is fixed to zero. 
Due to the limited size of the \BdtoKstarKs sample, the relative uncertainty obtained for the \BstoKstarKs decays is also applied to the \BdtoKstarKs decays.

Potential biases that may be associated with the maximum likelihood estimator are investigated using pseudoexperiments. 
The systematic uncertainty is determined as the average difference between the nominal value and the fitted yields in the pseudoexperiments.

The impact of the limited size of the simulated samples, used to determine the selection and particle identification efficiencies, is considered as systematic uncertainty.
In addition, the hardware trigger is a potential source of systematic uncertainty due to imperfections in the description of data by simulation.
A data sample of $D^{*+} \to D^0 \pi^{\pm}$, with \decay{\Dz}{K^- \pi^+}, decays is used to characterise the trigger efficiencies of the pions and kaons, separated
according to particle charge, as a function of the transverse energy of the associated cluster in the hadron calorimeter~\cite{LHCb-DP-2012-004,CERN-THESIS-2013-311}. 
These data-driven calibration curves are used to weight simulated events in order to determine the efficiency of the hadron trigger.

The effective lifetimes of \Bs meson reconstructed in a particular decay depend on the \CP-admixture of the final state
because \CP-even and \CP-odd eigenstates may have different lifetimes~\cite{Lifetime}.
Since the selection efficiency depends on decay time, this might lead to a source of uncertainty in the measurement.
The relative change in efficiency with respect to the nominal value, estimated for the extreme ranges of possible 
effective lifetime distributions, is assigned as the systematic uncertainty.

Finally, the uncertainty from the measurement of the fragmentation fractions ratio,
$f_s/f_d$~\cite{LHCb-PAPER-2011-018,LHCb-PAPER-2012-037,LHCb-CONF-2013-011}, is taken into account. 
A summary of the relative uncertainties on the ratio of branching fractions is given in Table~\ref{syst}.
The final results reported in Sec.~\ref{conclusion} take into account correlations between the two samples; thus the systematic uncertainty for the combined measurement is reduced.

\begin{table}[t]
    \caption{\small
    Systematic uncertainties on the relative branching fraction measurement for the two \KS categories. 
    The uncertainties are quoted as fractional contributions of the relative branching fraction and the total is the sum in quadrature of all contributions.
    }
\label{syst}
\begin{center}
\begin{tabular}{l c c c c}
\hline
Source & \multicolumn{2}{c}{$\frac{\BF(\BstoKstarKs)}{\BF(\BdtoKsPiPi)}$} & \multicolumn{2}{c}{$\frac{\BF(\BdtoKstarKs)}{\BF(\BdtoKsPiPi)}$} \\
\cline{2-5}
 & Downstream & Long & Downstream & Long \\
\hline
Fit                  &0.05  &0.03  &0.20  &0.28 \\
Selection efficiency       &0.08  &0.10  &0.08  &0.11 \\
PID efficiency               &0.01  &0.01  &0.01  &0.01 \\
Trigger           &0.07  &0.07  &0.02  &0.09 \\
Lifetime	         &0.05  &0.05  & -    & -   \\
\hline
Total               &0.13  &0.14  &0.22  &0.31 \\
\hline
$f_{s}/f_{d}$   & 0.06  & 0.06     & - & - \\
\hline
\end{tabular}
\end{center}
\end{table}

\section{Summary and conclusion}
\label{conclusion}
A search for \myBdstoKstarKs decays is performed by the LHCb experiment using $pp$ data recorded at a centre-of-mass energy of $7\tev$, corresponding to an integrated luminosity of $1.0$\invfb. 
The branching ratios 
are determined using the \BdtoKsPiPi decay as a normalisation mode.
The measurements are performed separately for the downstream and long \KS categories and then
combined following Refs.~\cite{Lyons1988110, Valassi2003391}. 

The \Bs decay is observed for the first time, with a total significance of $7.1$ standard deviations. 
The relative branching fraction is
\begin{eqnarray*}
\frac{\BF{(\BstoKstarKs)}}{\BF{(\BdtoKsPiPi)}} &=& 0.33 \pm 0.07 \ (\text{stat}) \pm 0.04 \ (\text{syst}) \pm 0.02 \ (f_s/f_d).
\end{eqnarray*}

For the \Bd decay, an upper limit at 90\% (95\%) confidence level (CL) is determined.
The likelihood function is convolved with a Gaussian function with standard deviation equal to the total systematic uncertainty,
and the upper limit is taken to be the value of the relative branching fraction below which 90\% (95\%) of the total integral
of the likelihood function over non-negative branching ratio values is found.
The central value and the upper limit on the relative branching fraction of the decay \BdtoKstarKs are
\begin{eqnarray*}
\frac{\BF{(\BdtoKstarKs)}}{\BF{(\BdtoKsPiPi)}} &=& 0.005 \pm 0.007 \ (\text{stat}) \pm 0.001 \ (\text{syst}), \\
                                               &<& 0.020\ (0.021) \; {\rm at} \; 90\% \;(95\%) \;\text{CL}. \\
\end{eqnarray*}

The absolute branching fractions, calculated using the reference value of 
$\BF(\BdtoKzPiPi) = (4.96 \pm 0.20) \times 10^{-5}$~\cite{PDG2012}, determined without using
the correlated LHCb measurement.  The results are expressed in terms of the sum of final states containing
either $K^0$ or $\bar{K}^0$ mesons
\begin{eqnarray*}
\mathcal{B}(B_{s}^{0} \to \bar{K}^{0} K^{*}(892)^{0}) + \mathcal{B}(B_{s}^{0} \to K^{0} \bar{K}^{*}(892)^{0}) &=&  (16.4 \pm 3.4  \pm 1.9 \pm 1.0 \pm 0.7) \times 10^{-6},\\
\mathcal{B}(B^{0} \to \bar{K}^{0} K^{*}(892)^{0}) + \mathcal{B}(B^{0} \to K^{0} \bar{K}^{*}(892)^{0}) &=&  (0.25 \pm 0.34 \pm 0.05 \pm 0.01) \times 10^{-6}, \\
                    &<&  0.96 \ (1.04) \times 10^{-6} \; {\rm at} \; 90\% \;(95\%) \;\text{CL},
\end{eqnarray*}
where the first uncertainty is statistical, the second systematic, the third due to the ratio of the fragmentation 
fractions and the fourth due to the uncertainty on the branching fraction of the normalisation decay. 
These results are in agreement with theoretical predictions~\cite{Cheng:2009mu,Ali:2007ff,Su:2011eq}
and can be used to further constrain phenomenological models.

\section*{Acknowledgements}

 
\noindent We express our gratitude to our colleagues in the CERN
accelerator departments for the excellent performance of the LHC. We
thank the technical and administrative staff at the LHCb
institutes. We acknowledge support from CERN and from the national
agencies: CAPES, CNPq, FAPERJ and FINEP (Brazil); NSFC (China);
CNRS/IN2P3 (France); BMBF, DFG, HGF and MPG (Germany); INFN (Italy); 
FOM and NWO (The Netherlands); MNiSW and NCN (Poland); MEN/IFA (Romania); 
MinES and FANO (Russia); MinECo (Spain); SNSF and SER (Switzerland); 
NASU (Ukraine); STFC (United Kingdom); NSF (USA).
The Tier1 computing centres are supported by IN2P3 (France), KIT and BMBF 
(Germany), INFN (Italy), NWO and SURF (The Netherlands), PIC (Spain), GridPP 
(United Kingdom).
We are indebted to the communities behind the multiple open 
source software packages on which we depend. We are also thankful for the 
computing resources and the access to software R\&D tools provided by Yandex LLC (Russia).
Individual groups or members have received support from 
EPLANET, Marie Sk\l{}odowska-Curie Actions and ERC (European Union), 
Conseil g\'{e}n\'{e}ral de Haute-Savoie, Labex ENIGMASS and OCEVU, 
R\'{e}gion Auvergne (France), RFBR (Russia), XuntaGal and GENCAT (Spain), Royal Society and Royal
Commission for the Exhibition of 1851 (United Kingdom).
We acknowledge Ulrich Nierste from KIT (Germany) for his assistance on theoretical aspects of the analysis.



\addcontentsline{toc}{section}{References}
\setboolean{inbibliography}{true}
\bibliographystyle{LHCb}
\bibliography{main,LHCb-PAPER,LHCb-CONF,LHCb-DP,LHCb-TDR}

\ifx\mcitethebibliography\mciteundefinedmacro
\PackageError{LHCb.bst}{mciteplus.sty has not been loaded}
{This bibstyle requires the use of the mciteplus package.}\fi
\providecommand{\href}[2]{#2}
\begin{mcitethebibliography}{10}
\mciteSetBstSublistMode{n}
\mciteSetBstMaxWidthForm{subitem}{\alph{mcitesubitemcount})}
\mciteSetBstSublistLabelBeginEnd{\mcitemaxwidthsubitemform\space}
{\relax}{\relax}

\bibitem{sakharov}
A.~D. Sakharov, \ifthenelse{\boolean{articletitles}}{\emph{{Violation of \CP
  invariance, C asymmetry, and baryon asymmetry of the Universe}},
  }{}\href{http://dx.doi.org/10.1070/PU1991v034n05ABEH002497}{J.\ Exp.\ Theor.\
  Phys.\ Lett.\  \textbf{5} (1967) 24}\relax
\mciteBstWouldAddEndPuncttrue
\mciteSetBstMidEndSepPunct{\mcitedefaultmidpunct}
{\mcitedefaultendpunct}{\mcitedefaultseppunct}\relax
\EndOfBibitem
\bibitem{Cabibbo}
N.~Cabibbo, \ifthenelse{\boolean{articletitles}}{\emph{{Unitary symmetry and
  leptonic decays}},
  }{}\href{http://dx.doi.org/10.1103/PhysRevLett.10.531}{Phys.\ Rev.\ Lett.\
  \textbf{10} (1963) 531}\relax
\mciteBstWouldAddEndPuncttrue
\mciteSetBstMidEndSepPunct{\mcitedefaultmidpunct}
{\mcitedefaultendpunct}{\mcitedefaultseppunct}\relax
\EndOfBibitem
\bibitem{Kobayashi}
M.~Kobayashi and T.~Maskawa, \ifthenelse{\boolean{articletitles}}{\emph{{\CP
  violation in the renormalizable theory of weak interaction}},
  }{}\href{http://dx.doi.org/10.1143/PTP.49.652}{Prog.\ Theor.\ Phys.\
  \textbf{49} (1973) 652}\relax
\mciteBstWouldAddEndPuncttrue
\mciteSetBstMidEndSepPunct{\mcitedefaultmidpunct}
{\mcitedefaultendpunct}{\mcitedefaultseppunct}\relax
\EndOfBibitem
\bibitem{Riotto}
A.~Riotto and M.~Trodden, \ifthenelse{\boolean{articletitles}}{\emph{{Recent
  progress in baryogenesis}},
  }{}\href{http://dx.doi.org/10.1146/annurev.nucl.49.1.35}{Annu.\ Rev.\ Nucl.\
  Part.\ Sci.\  \textbf{49} (1999) 35},
  \href{http://arxiv.org/abs/hep-ph/9901362}{{\tt arXiv:hep-ph/9901362}}\relax
\mciteBstWouldAddEndPuncttrue
\mciteSetBstMidEndSepPunct{\mcitedefaultmidpunct}
{\mcitedefaultendpunct}{\mcitedefaultseppunct}\relax
\EndOfBibitem
\bibitem{Ciuchini}
M.~Ciuchini, M.~Pierini, and L.~Silvestrini,
  \ifthenelse{\boolean{articletitles}}{\emph{{$B_{s}^{0} \to K^{(*)0}
  \overline{K}^{(*)0}$ \CP asymmetries: golden channels for New Physics
  Searches}}, }{}\href{http://dx.doi.org/10.1103/PhysRevLett.100.031802}{Phys.\
  Rev.\ Lett.\  \textbf{100} (2008) 031802}\relax
\mciteBstWouldAddEndPuncttrue
\mciteSetBstMidEndSepPunct{\mcitedefaultmidpunct}
{\mcitedefaultendpunct}{\mcitedefaultseppunct}\relax
\EndOfBibitem
\bibitem{Cheng:2009mu}
H.-Y. Cheng and C.-K. Chua, \ifthenelse{\boolean{articletitles}}{\emph{{QCD
  factorization for charmless hadronic \Bs decays revisited}},
  }{}\href{http://dx.doi.org/10.1103/PhysRevD.80.114026}{Phys.\ Rev.\
  \textbf{D80} (2009) 114026}, \href{http://arxiv.org/abs/0910.5237}{{\tt
  arXiv:0910.5237}}\relax
\mciteBstWouldAddEndPuncttrue
\mciteSetBstMidEndSepPunct{\mcitedefaultmidpunct}
{\mcitedefaultendpunct}{\mcitedefaultseppunct}\relax
\EndOfBibitem
\bibitem{Ali:2007ff}
A.~Ali {\em et~al.}, \ifthenelse{\boolean{articletitles}}{\emph{{Charmless
  nonleptonic $\Bs$ decays to $PP$, $PV$, and $VV$ final states in the
  perturbative QCD approach}},
  }{}\href{http://dx.doi.org/10.1103/PhysRevD.76.074018}{Phys.\ Rev.\
  \textbf{D76} (2007) 074018}, \href{http://arxiv.org/abs/hep-ph/0703162}{{\tt
  arXiv:hep-ph/0703162}}\relax
\mciteBstWouldAddEndPuncttrue
\mciteSetBstMidEndSepPunct{\mcitedefaultmidpunct}
{\mcitedefaultendpunct}{\mcitedefaultseppunct}\relax
\EndOfBibitem
\bibitem{Su:2011eq}
F.~Su, Y.-L. Wu, C.~Zhuang, and Y.-B. Yang,
  \ifthenelse{\boolean{articletitles}}{\emph{{Charmless $\Bs\to PP, PV, VV$
  decays based on the six-quark effective Hamiltonian with strong phase effects
  II}}, }{}\href{http://dx.doi.org/10.1140/epjc/s10052-012-1914-4}{Eur.\ Phys.\
  J.\  \textbf{C72} (2012) 1914}, \href{http://arxiv.org/abs/1107.0136}{{\tt
  arXiv:1107.0136}}\relax
\mciteBstWouldAddEndPuncttrue
\mciteSetBstMidEndSepPunct{\mcitedefaultmidpunct}
{\mcitedefaultendpunct}{\mcitedefaultseppunct}\relax
\EndOfBibitem
\bibitem{Dalitz:1953cp}
R.~H. Dalitz, \ifthenelse{\boolean{articletitles}}{\emph{{On the analysis of
  $\tau$-meson data and the nature of the $\tau$-meson}},
  }{}\href{http://dx.doi.org/10.1080/14786441008520365}{Phil.\ Mag.\
  \textbf{44} (1953) 1068}\relax
\mciteBstWouldAddEndPuncttrue
\mciteSetBstMidEndSepPunct{\mcitedefaultmidpunct}
{\mcitedefaultendpunct}{\mcitedefaultseppunct}\relax
\EndOfBibitem
\bibitem{LHCb-PAPER-2013-042}
LHCb collaboration, R.~Aaij {\em et~al.},
  \ifthenelse{\boolean{articletitles}}{\emph{{Study of $B_{(s)}^0 \to K^0_S h^+
  h^{\prime -}$ decays with first observation of $B_s^0 \to K^0_S K^\pm
  \pi^\mp$ and $B_s^0 \to K^0_S \pi^+\pi^-$}},
  }{}\href{http://dx.doi.org/10.1007/JHEP10(2013)143}{JHEP \textbf{10} (2013)
  143}, \href{http://arxiv.org/abs/1307.7648}{{\tt arXiv:1307.7648}}\relax
\mciteBstWouldAddEndPuncttrue
\mciteSetBstMidEndSepPunct{\mcitedefaultmidpunct}
{\mcitedefaultendpunct}{\mcitedefaultseppunct}\relax
\EndOfBibitem
\bibitem{LHCb-PAPER-2014-043}
LHCb collaboration, R.~Aaij {\em et~al.},
  \ifthenelse{\boolean{articletitles}}{\emph{{Observation of $B^0_s \to
  K^{*\pm}K^\mp$ and evidence of $B^0_s \to K^{*-}\pi^+$ decays}},
  }{}\href{http://dx.doi.org/10.1088/1367-2630/16/12/123001}{New J.\ Phys.\
  \textbf{16} (2014) 123001}, \href{http://arxiv.org/abs/1407.7704}{{\tt
  arXiv:1407.7704}}\relax
\mciteBstWouldAddEndPuncttrue
\mciteSetBstMidEndSepPunct{\mcitedefaultmidpunct}
{\mcitedefaultendpunct}{\mcitedefaultseppunct}\relax
\EndOfBibitem
\bibitem{Aubert:2006wu}
\babar collaboration, B.~Aubert {\em et~al.},
  \ifthenelse{\boolean{articletitles}}{\emph{{Search for the decay of a \Bd or
  \Bdb meson to $\Kstarzb\Kz$ or $\Kstarz\Kzb$}},
  }{}\href{http://dx.doi.org/10.1103/PhysRevD.74.072008}{Phys.\ Rev.\
  \textbf{D74} (2006) 072008}, \href{http://arxiv.org/abs/hep-ex/0606050}{{\tt
  arXiv:hep-ex/0606050}}\relax
\mciteBstWouldAddEndPuncttrue
\mciteSetBstMidEndSepPunct{\mcitedefaultmidpunct}
{\mcitedefaultendpunct}{\mcitedefaultseppunct}\relax
\EndOfBibitem
\bibitem{PDG2014}
Particle Data Group, K.~A. Olive {\em et~al.},
  \ifthenelse{\boolean{articletitles}}{\emph{{\href{http://pdg.lbl.gov/}{Review
  of particle physics}}},
  }{}\href{http://dx.doi.org/10.1088/1674-1137/38/9/090001}{Chin.\ Phys.\
  \textbf{C38} (2014) 090001}\relax
\mciteBstWouldAddEndPuncttrue
\mciteSetBstMidEndSepPunct{\mcitedefaultmidpunct}
{\mcitedefaultendpunct}{\mcitedefaultseppunct}\relax
\EndOfBibitem
\bibitem{LHCb-PAPER-2011-018}
LHCb collaboration, R.~Aaij {\em et~al.},
  \ifthenelse{\boolean{articletitles}}{\emph{{Measurement of $b$ hadron
  production fractions in 7 TeV $pp$ collisions}},
  }{}\href{http://dx.doi.org/10.1103/PhysRevD.85.032008}{Phys.\ Rev.\
  \textbf{D85} (2012) 032008}, \href{http://arxiv.org/abs/1111.2357}{{\tt
  arXiv:1111.2357}}\relax
\mciteBstWouldAddEndPuncttrue
\mciteSetBstMidEndSepPunct{\mcitedefaultmidpunct}
{\mcitedefaultendpunct}{\mcitedefaultseppunct}\relax
\EndOfBibitem
\bibitem{LHCb-PAPER-2012-037}
LHCb collaboration, R.~Aaij {\em et~al.},
  \ifthenelse{\boolean{articletitles}}{\emph{{Measurement of the fragmentation
  fraction ratio $f_s/f_d$ and its dependence on $B$ meson kinematics}},
  }{}\href{http://dx.doi.org/10.1007/JHEP04(2013)001}{JHEP \textbf{04} (2013)
  001}, \href{http://arxiv.org/abs/1301.5286}{{\tt arXiv:1301.5286}}\relax
\mciteBstWouldAddEndPuncttrue
\mciteSetBstMidEndSepPunct{\mcitedefaultmidpunct}
{\mcitedefaultendpunct}{\mcitedefaultseppunct}\relax
\EndOfBibitem
\bibitem{LHCb-CONF-2013-011}
{LHCb collaboration}, \ifthenelse{\boolean{articletitles}}{\emph{{Updated
  average $f_{s}/f_{d}$ $b$-hadron production fraction ratio for $7 \tev$ $pp$
  collisions}}, }{}
  \href{http://cdsweb.cern.ch/search?p=LHCb-CONF-2013-011&f=reportnumber&action_search=Search&c=LHCb+Conference+Contributions}
  {LHCb-CONF-2013-011}\relax
\mciteBstWouldAddEndPuncttrue
\mciteSetBstMidEndSepPunct{\mcitedefaultmidpunct}
{\mcitedefaultendpunct}{\mcitedefaultseppunct}\relax
\EndOfBibitem
\bibitem{Alves:2008zz}
LHCb collaboration, A.~A. Alves~Jr.\ {\em et~al.},
  \ifthenelse{\boolean{articletitles}}{\emph{{The \lhcb detector at the LHC}},
  }{}\href{http://dx.doi.org/10.1088/1748-0221/3/08/S08005}{JINST \textbf{3}
  (2008) S08005}\relax
\mciteBstWouldAddEndPuncttrue
\mciteSetBstMidEndSepPunct{\mcitedefaultmidpunct}
{\mcitedefaultendpunct}{\mcitedefaultseppunct}\relax
\EndOfBibitem
\bibitem{LHCb-DP-2014-002}
LHCb collaboration, R.~Aaij {\em et~al.},
  \ifthenelse{\boolean{articletitles}}{\emph{{LHCb detector performance}},
  }{}\href{http://dx.doi.org/10.1142/S0217751X15300227}{Int.\ J.\ Mod.\ Phys.\
  \textbf{A30} (2015) 1530022}, \href{http://arxiv.org/abs/1412.6352}{{\tt
  arXiv:1412.6352}}\relax
\mciteBstWouldAddEndPuncttrue
\mciteSetBstMidEndSepPunct{\mcitedefaultmidpunct}
{\mcitedefaultendpunct}{\mcitedefaultseppunct}\relax
\EndOfBibitem
\bibitem{Sjostrand:2006za}
T.~Sj\"{o}strand, S.~Mrenna, and P.~Skands,
  \ifthenelse{\boolean{articletitles}}{\emph{{PYTHIA 6.4 physics and manual}},
  }{}\href{http://dx.doi.org/10.1088/1126-6708/2006/05/026}{JHEP \textbf{05}
  (2006) 026}, \href{http://arxiv.org/abs/hep-ph/0603175}{{\tt
  arXiv:hep-ph/0603175}}\relax
\mciteBstWouldAddEndPuncttrue
\mciteSetBstMidEndSepPunct{\mcitedefaultmidpunct}
{\mcitedefaultendpunct}{\mcitedefaultseppunct}\relax
\EndOfBibitem
\bibitem{LHCb-PROC-2010-056}
I.~Belyaev {\em et~al.}, \ifthenelse{\boolean{articletitles}}{\emph{{Handling
  of the generation of primary events in Gauss, the LHCb simulation
  framework}}, }{}\href{http://dx.doi.org/10.1088/1742-6596/331/3/032047}{{J.\
  Phys.\ Conf.\ Ser.\ } \textbf{331} (2011) 032047}\relax
\mciteBstWouldAddEndPuncttrue
\mciteSetBstMidEndSepPunct{\mcitedefaultmidpunct}
{\mcitedefaultendpunct}{\mcitedefaultseppunct}\relax
\EndOfBibitem
\bibitem{Lange:2001uf}
D.~J. Lange, \ifthenelse{\boolean{articletitles}}{\emph{{The EvtGen particle
  decay simulation package}},
  }{}\href{http://dx.doi.org/10.1016/S0168-9002(01)00089-4}{Nucl.\ Instrum.\
  Meth.\  \textbf{A462} (2001) 152}\relax
\mciteBstWouldAddEndPuncttrue
\mciteSetBstMidEndSepPunct{\mcitedefaultmidpunct}
{\mcitedefaultendpunct}{\mcitedefaultseppunct}\relax
\EndOfBibitem
\bibitem{Golonka:2005pn}
P.~Golonka and Z.~Was, \ifthenelse{\boolean{articletitles}}{\emph{{PHOTOS Monte
  Carlo: A precision tool for QED corrections in $Z$ and $W$ decays}},
  }{}\href{http://dx.doi.org/10.1140/epjc/s2005-02396-4}{Eur.\ Phys.\ J.\
  \textbf{C45} (2006) 97}, \href{http://arxiv.org/abs/hep-ph/0506026}{{\tt
  arXiv:hep-ph/0506026}}\relax
\mciteBstWouldAddEndPuncttrue
\mciteSetBstMidEndSepPunct{\mcitedefaultmidpunct}
{\mcitedefaultendpunct}{\mcitedefaultseppunct}\relax
\EndOfBibitem
\bibitem{Allison:2006ve}
Geant4 collaboration, J.~Allison {\em et~al.},
  \ifthenelse{\boolean{articletitles}}{\emph{{Geant4 developments and
  applications}}, }{}\href{http://dx.doi.org/10.1109/TNS.2006.869826}{IEEE
  Trans.\ Nucl.\ Sci.\  \textbf{53} (2006) 270}\relax
\mciteBstWouldAddEndPuncttrue
\mciteSetBstMidEndSepPunct{\mcitedefaultmidpunct}
{\mcitedefaultendpunct}{\mcitedefaultseppunct}\relax
\EndOfBibitem
\bibitem{Agostinelli:2002hh}
Geant4 collaboration, S.~Agostinelli {\em et~al.},
  \ifthenelse{\boolean{articletitles}}{\emph{{Geant4: a simulation toolkit}},
  }{}\href{http://dx.doi.org/10.1016/S0168-9002(03)01368-8}{Nucl.\ Instrum.\
  Meth.\  \textbf{A506} (2003) 250}\relax
\mciteBstWouldAddEndPuncttrue
\mciteSetBstMidEndSepPunct{\mcitedefaultmidpunct}
{\mcitedefaultendpunct}{\mcitedefaultseppunct}\relax
\EndOfBibitem
\bibitem{LHCb-PROC-2011-006}
M.~Clemencic {\em et~al.}, \ifthenelse{\boolean{articletitles}}{\emph{{The
  \lhcb simulation application, Gauss: Design, evolution and experience}},
  }{}\href{http://dx.doi.org/10.1088/1742-6596/331/3/032023}{{J.\ Phys.\ Conf.\
  Ser.\ } \textbf{331} (2011) 032023}\relax
\mciteBstWouldAddEndPuncttrue
\mciteSetBstMidEndSepPunct{\mcitedefaultmidpunct}
{\mcitedefaultendpunct}{\mcitedefaultseppunct}\relax
\EndOfBibitem
\bibitem{LHCb-DP-2012-004}
R.~Aaij {\em et~al.}, \ifthenelse{\boolean{articletitles}}{\emph{{The \lhcb
  trigger and its performance in 2011}},
  }{}\href{http://dx.doi.org/10.1088/1748-0221/8/04/P04022}{JINST \textbf{8}
  (2013) P04022}, \href{http://arxiv.org/abs/1211.3055}{{\tt
  arXiv:1211.3055}}\relax
\mciteBstWouldAddEndPuncttrue
\mciteSetBstMidEndSepPunct{\mcitedefaultmidpunct}
{\mcitedefaultendpunct}{\mcitedefaultseppunct}\relax
\EndOfBibitem
\bibitem{BBDT}
V.~V. Gligorov and M.~Williams,
  \ifthenelse{\boolean{articletitles}}{\emph{{Efficient, reliable and fast
  high-level triggering using a bonsai boosted decision tree}},
  }{}\href{http://dx.doi.org/10.1088/1748-0221/8/02/P02013}{JINST \textbf{8}
  (2013) P02013}, \href{http://arxiv.org/abs/1210.6861}{{\tt
  arXiv:1210.6861}}\relax
\mciteBstWouldAddEndPuncttrue
\mciteSetBstMidEndSepPunct{\mcitedefaultmidpunct}
{\mcitedefaultendpunct}{\mcitedefaultseppunct}\relax
\EndOfBibitem
\bibitem{Breiman}
L.~Breiman, J.~H. Friedman, R.~A. Olshen, and C.~J. Stone, {\em Classification
  and regression trees}, Wadsworth international group, Belmont, California,
  USA, 1984\relax
\mciteBstWouldAddEndPuncttrue
\mciteSetBstMidEndSepPunct{\mcitedefaultmidpunct}
{\mcitedefaultendpunct}{\mcitedefaultseppunct}\relax
\EndOfBibitem
\bibitem{AdaBoost}
R.~E. Schapire and Y.~Freund, \ifthenelse{\boolean{articletitles}}{\emph{A
  decision-theoretic generalization of on-line learning and an application to
  boosting}, }{}\href{http://dx.doi.org/10.1006/jcss.1997.1504}{Jour.\ Comp.\
  and Syst.\ Sc.\  \textbf{55} (1997) 119}\relax
\mciteBstWouldAddEndPuncttrue
\mciteSetBstMidEndSepPunct{\mcitedefaultmidpunct}
{\mcitedefaultendpunct}{\mcitedefaultseppunct}\relax
\EndOfBibitem
\bibitem{Punzi:2003bu}
G.~Punzi, \ifthenelse{\boolean{articletitles}}{\emph{{Sensitivity of searches
  for new signals and its optimization}}, }{} in {\em Statistical Problems in
  Particle Physics, Astrophysics, and Cosmology} (L.~{Lyons}, R.~{Mount}, and
  R.~{Reitmeyer}, eds.), p.~79, 2003.
\newblock \href{http://arxiv.org/abs/physics/0308063}{{\tt
  arXiv:physics/0308063}}\relax
\mciteBstWouldAddEndPuncttrue
\mciteSetBstMidEndSepPunct{\mcitedefaultmidpunct}
{\mcitedefaultendpunct}{\mcitedefaultseppunct}\relax
\EndOfBibitem
\bibitem{Pivk:2004ty}
M.~Pivk and F.~R. Le~Diberder,
  \ifthenelse{\boolean{articletitles}}{\emph{{sPlot: A statistical tool to
  unfold data distributions}},
  }{}\href{http://dx.doi.org/10.1016/j.nima.2005.08.106}{Nucl.\ Instrum.\
  Meth.\  \textbf{A555} (2005) 356},
  \href{http://arxiv.org/abs/physics/0402083}{{\tt
  arXiv:physics/0402083}}\relax
\mciteBstWouldAddEndPuncttrue
\mciteSetBstMidEndSepPunct{\mcitedefaultmidpunct}
{\mcitedefaultendpunct}{\mcitedefaultseppunct}\relax
\EndOfBibitem
\bibitem{Wilks:1938dza}
S.~S. Wilks, \ifthenelse{\boolean{articletitles}}{\emph{{The large-sample
  distribution of the likelihood ratio for testing composite hypotheses}},
  }{}\href{http://dx.doi.org/10.1214/aoms/1177732360}{Ann.\ Math.\ Statist.\
  \textbf{9} (1938) 60}\relax
\mciteBstWouldAddEndPuncttrue
\mciteSetBstMidEndSepPunct{\mcitedefaultmidpunct}
{\mcitedefaultendpunct}{\mcitedefaultseppunct}\relax
\EndOfBibitem
\bibitem{CERN-THESIS-2013-311}
{A.\ Martin Sanchez}, {\em {CP violation studies on the $B^0 \to D K^{*0}$
  decays and hadronic trigger performance with the LHCb detector at CERN}}, PhD
  thesis, Universit\'e Paris-sud XI, {2013},
  \href{http://cds.cern.ch/record/1599138}{CERN-THESIS-2013-311}\relax
\mciteBstWouldAddEndPuncttrue
\mciteSetBstMidEndSepPunct{\mcitedefaultmidpunct}
{\mcitedefaultendpunct}{\mcitedefaultseppunct}\relax
\EndOfBibitem
\bibitem{Lifetime}
I.~Dunietz, R.~Fleischer, and U.~Nierste,
  \ifthenelse{\boolean{articletitles}}{\emph{{In pursuit of new physics with
  $\Bs$ decays}}, }{}\href{http://dx.doi.org/10.1103/PhysRevD.63.114015}{Phys.\
  Rev.\ D \textbf{63} (2001) 114015}, \href{http://arxiv.org/abs/0012219}{{\tt
  arXiv:0012219}}\relax
\mciteBstWouldAddEndPuncttrue
\mciteSetBstMidEndSepPunct{\mcitedefaultmidpunct}
{\mcitedefaultendpunct}{\mcitedefaultseppunct}\relax
\EndOfBibitem
\bibitem{Lyons1988110}
L.~Lyons, D.~Gibaut, and P.~Clifford,
  \ifthenelse{\boolean{articletitles}}{\emph{How to combine correlated
  estimates of a single physical quantity},
  }{}\href{http://dx.doi.org/http://dx.doi.org/10.1016/0168-9002(88)90018-6}{Nucl.\
  Instrum.\ Meth.\  \textbf{A270} (1988) 110}\relax
\mciteBstWouldAddEndPuncttrue
\mciteSetBstMidEndSepPunct{\mcitedefaultmidpunct}
{\mcitedefaultendpunct}{\mcitedefaultseppunct}\relax
\EndOfBibitem
\bibitem{Valassi2003391}
A.~Valassi, \ifthenelse{\boolean{articletitles}}{\emph{Combining correlated
  measurements of several different physical quantities},
  }{}\href{http://dx.doi.org/http://dx.doi.org/10.1016/S0168-9002(03)00329-2}{Nucl.\
  Instrum.\ Meth.\  \textbf{A500} (2003) 391}\relax
\mciteBstWouldAddEndPuncttrue
\mciteSetBstMidEndSepPunct{\mcitedefaultmidpunct}
{\mcitedefaultendpunct}{\mcitedefaultseppunct}\relax
\EndOfBibitem
\bibitem{PDG2012}
Particle Data Group, J.~Beringer {\em et~al.},
  \ifthenelse{\boolean{articletitles}}{\emph{{\href{http://pdg.lbl.gov/}{Review
  of particle physics}}},
  }{}\href{http://dx.doi.org/10.1103/PhysRevD.86.010001}{Phys.\ Rev.\
  \textbf{D86} (2012) 010001}\relax
\mciteBstWouldAddEndPuncttrue
\mciteSetBstMidEndSepPunct{\mcitedefaultmidpunct}
{\mcitedefaultendpunct}{\mcitedefaultseppunct}\relax
\EndOfBibitem
\end{mcitethebibliography}

\newpage

\newpage


\centerline{\large\bf LHCb collaboration}
\begin{flushleft}
\small
R.~Aaij$^{38}$, 
B.~Adeva$^{37}$, 
M.~Adinolfi$^{46}$, 
A.~Affolder$^{52}$, 
Z.~Ajaltouni$^{5}$, 
S.~Akar$^{6}$, 
J.~Albrecht$^{9}$, 
F.~Alessio$^{38}$, 
M.~Alexander$^{51}$, 
S.~Ali$^{41}$
% ...

{\footnotesize \it
% ...
$ ^{5}$Clermont Universit\'{e}, Universit\'{e} Blaise Pascal, CNRS/IN2P3, LPC, Clermont-Ferrand, France\\
$ ^{6}$CPPM, Aix-Marseille Universit\'{e}, CNRS/IN2P3, Marseille, France\\
% ...
$ ^{9}$Fakult\"{a}t Physik, Technische Universit\"{a}t Dortmund, Dortmund, Germany\\
% ...
$ ^{37}$Universidad de Santiago de Compostela, Santiago de Compostela, Spain\\
$ ^{38}$European Organization for Nuclear Research (CERN), Geneva, Switzerland\\
% ...
$ ^{41}$Nikhef National Institute for Subatomic Physics, Amsterdam, The Netherlands\\
% ...
$ ^{46}$H.H. Wills Physics Laboratory, University of Bristol, Bristol, United Kingdom\\
% ...
$ ^{51}$School of Physics and Astronomy, University of Glasgow, Glasgow, United Kingdom\\
$ ^{52}$Oliver Lodge Laboratory, University of Liverpool, Liverpool, United Kingdom\\
% ...
}
\end{flushleft}

\newpage

%
%

\end{document}